# Nonlinear electron scattering activated by surface plasmon excitation of Ag nanostructures

ChunKai Xu[1], WenJie Liu[1], PanKe Zhang[1], KeZun Xu[1], Yi Luo[2,3,a] and XiangJun Chen[1,b]

[1]*Hefei National Lab for Physical Science at Microscale and Department of Modern Physics, University of Science and Technology of China, Hefei, 230026, China*

[2] *Hefei National Lab for Physical Science at Microscale and Department of Chemical Physics, University of Science and Technology of China, Hefei, 230026, China*

[3]*Department of Theoretical Chemistry, Royal Institute of Technology, S-106 91 Stockholm, Sweden*

The discovery of many fascinating new phenomena associated with the surface plasmon polariton (SPP) has triggered the rapid development of nanophotonics and nanoelectronics. We report here the experimental observation of a fundamentally new physical process, nonlinear electron scattering, stimulated by the SPP excitation of Ag nanostructures on graphite surface in scanning probe electron energy loss spectroscopy. The observed intensity of SPP energy loss peak normalized to the elastic scattering intensity shows clearly a quadratic dependence on the external electric field strength generated by the tip-sample bias. The strong coherent nature of the SPP has made the observation possible and a two-step scattering process is proposed to explain this novel nonlinear effect. Our findings shed new light on the nature of SPP and pave the way to new spectroscopic applications.

Surface plasmon polariton (SPP) has been the central focuses of several advanced new technologies, including nanophotonics, nanoelectronics and ultrasensitive spectroscopy [1-8]. It is well known that the surface plasmon can be generated by two different sources, light or electron excitations. The nature of the light-induced plasmon has been extensively studied using a variety of experimental and theoretical tools. On the other hand, although the surface plasmon excited by inelastic scattering of electrons has long been investigated by electron energy loss spectroscopy (EELS) performed either with high-energy electrons transmitted through the thin film or low-energy electron reflected from the surface [4,5,9-11], the nature of the excitation process has not been fully explored. In the present work, we have carefully examined the SPP of Ag nanostructures on graphite surface by scanning probe electron energy loss spectroscopy (SP-EELS) [12]. A fundamental new physical process, namely the nonlinear electron scattering, is observed. It is found that the intensity of SPP energy loss peak normalized to the elastic scattering intensity shows clearly a quadratic dependence on the external electric field strength generated by the tip-sample bias. Such behavior could be understood by a one-electron two-step scattering mechanism. The observation of nonlinear electron scattering also indicates that SPP activated by electron excitations is a coherent state.

---

[a] e-mail: yiluo@ustc.edu.cn

[b] e-mail: xjun@ustc.edu.cn

Nonlinear interaction between the photon and the matter is observable benefited from the high intensity and good coherence of the lasers. This could be well understood from a simple perturbation picture, $I=\alpha I_0+\beta I_0^2$, in which $I_0$ and $I$ are the intensity of the incoming and outgoing photon, $\alpha$ and $\beta$ are the linear and non-linear coefficients related to properties of the system. However, it is generally believed that the nonlinear interaction between the electron and the matter is very difficulty to be observed, if not impossible. It is because that both the electron density and the electric field in an electron beam are often too weak in comparison with the laser. One can also notice from the perturbation expression that the exceptional large nonlinear coefficient $\beta$ could also significantly enhance the contribution from the nonlinear term. In other words, the basic ingredients for the observation of nonlinear electron scattering are to increase the local electric field and to find states with large nonlinear coefficients. The former can be partially utilized with the scanning probe electron energy spectrometer [12] by adjusting the tip-sample distance and external voltage. The latter requirement could be met by the surface plasmon polaritons (SPP), which is highly polarizable by its nature. Moreover, it is well known that local electric field can be effectively enhanced by the presence of the plasmon [6,13,14]. With the strategy in mind, we design an experiment to study nonlinear electron scattering in SPP of the Ag nanostructures on graphite surface by scanning probe electron energy loss spectroscopy (SP-EELS).

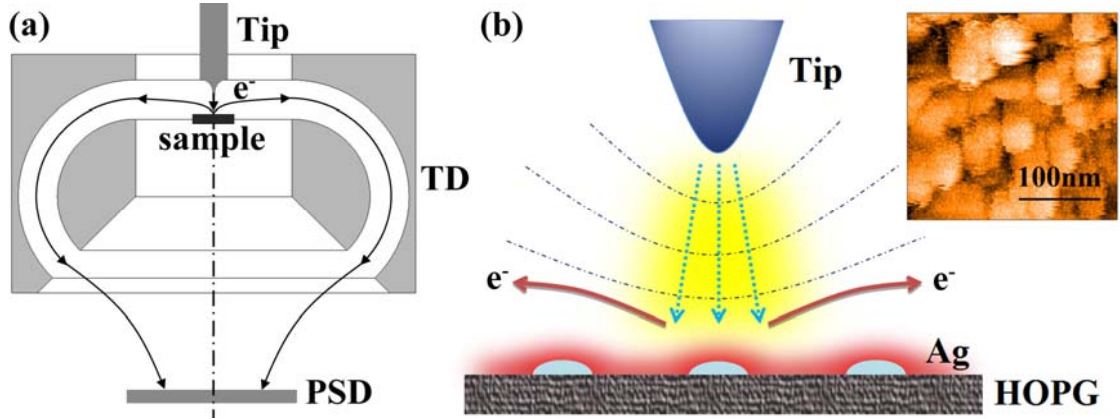

Figure 1: (a) Schematic drawing of the scanning probe electron energy spectrometer, which consists of a tip-sample system and a toroidal electron energy analyzer (TEEA). The backscattered electrons pass through a toroidal deflector (TD) and are detected by a two-dimensional position sensitive detector (PSD). (b) The possible electron scattering process involving in the SPP excitations. The inset shows the STM image of the sample surface.

Figure 1a presents the schematic drawing of the scanning probe electron energy spectrometer used in this experiment. The details can be found elsewhere [12]. Briefly, it consists of a tip-sample system and a toroidal electron energy analyzer (TEEA). A tip made from a 0.42 mm tungsten wire by electrochemical etching is approached to a distance of micrometers from the sample surface. Electrons are field emitted from the tip when a negative voltage $V_t$ of hundred volts is applied to the tip, while the sample is grounded. The backscattered electrons from the sample surface pass through a toroidal deflector and are detected by a two-dimensional position sensitive detector (PSD). The basic electron scattering process investigated in this work is sketched in Figure 1b. The sample is prepared by evaporating 35nm thin film of Ag on freshly cleaved HOPG. Ag structures with the dimension of tens of nanometers are observed on sample surface, as illustrated in the inset of Figure 1b. The SPP mode of the Ag nanostructures is excited by the incident electrons under a strong electric field introduced by the tip-sample bias and the electron energy loss spectrum (EELS) of the scattered electrons is acquired by TEEA.

The energy loss feature of Ag has been studied extensively by EELS including low energy backscattering EELS [10,15] and high energy transmission EELS [5,11]. A typical EELS spectrum of Ag nanostructures is shown in Figure 2, which is obtained at tip voltage -246V and sample current 10pA. An energy loss peak located at about 3.7 eV is known to be associated with the SPP of Ag. Our spectrum is in excellent agreement with the ones reported by Palmer's group, who also used the scanning probe electron spectroscopy to investigate the EELS of Ag surface [16,17]. In their experiments, a tip withdrawn a distance from the tunneling region was invoked as a field emission electron source when a bias voltage of hundred volts was applied. The energy of backscattered electrons was analyzed by a hemispherical deflector. One can see from these spectra, the Ag SPP excitation peak is much weaker than the main elastic peak.

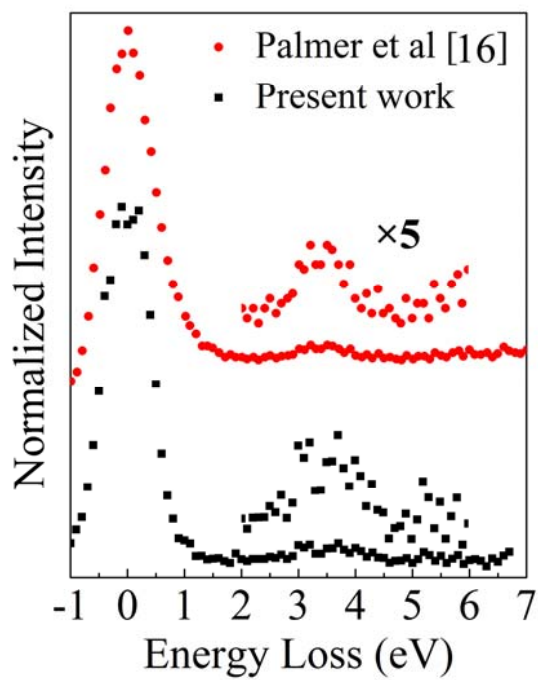

Figure 2: EELS acquired at the tip voltage -246V and the sample current 10pA (black solid square), which are compared with the one obtained at the tip voltage -170V and the sample current 10nA by Palmer's group (red solid circle) [16]. The spectral features related to the plasmon states are amplified by a factor 5.

By increasing the tip voltage at the fixed tip-sample distance and carrying out the same measurement, EELS at different electric field are obtained, five of which are illustrated in Figure 3a. The spectra have been background-subtracted by polynomial function [17], normalized by the intensity of the elastic scattering (ES) peak, and shifted vertically on the y axis for a better comparison. Gaussian function fitted curves of the ES peaks and SPP peaks are also shown as solid line. It can be seen that the increase of the electric field has significantly effects on the intensity of the SPP peak, which is enhanced drastically when the value of $V_t$ goes beyond 250V. This observation provides unambiguous evidence that the nonlinear electron scattering is a measurable process for the SPP. To better demonstrate this nonlinear effect, the relative intensity (*RI*) of the SPP peak, which is the area ratio of the SPP peak to the ES peak, is plotted as a function of $V_t$ in Figure 3b.

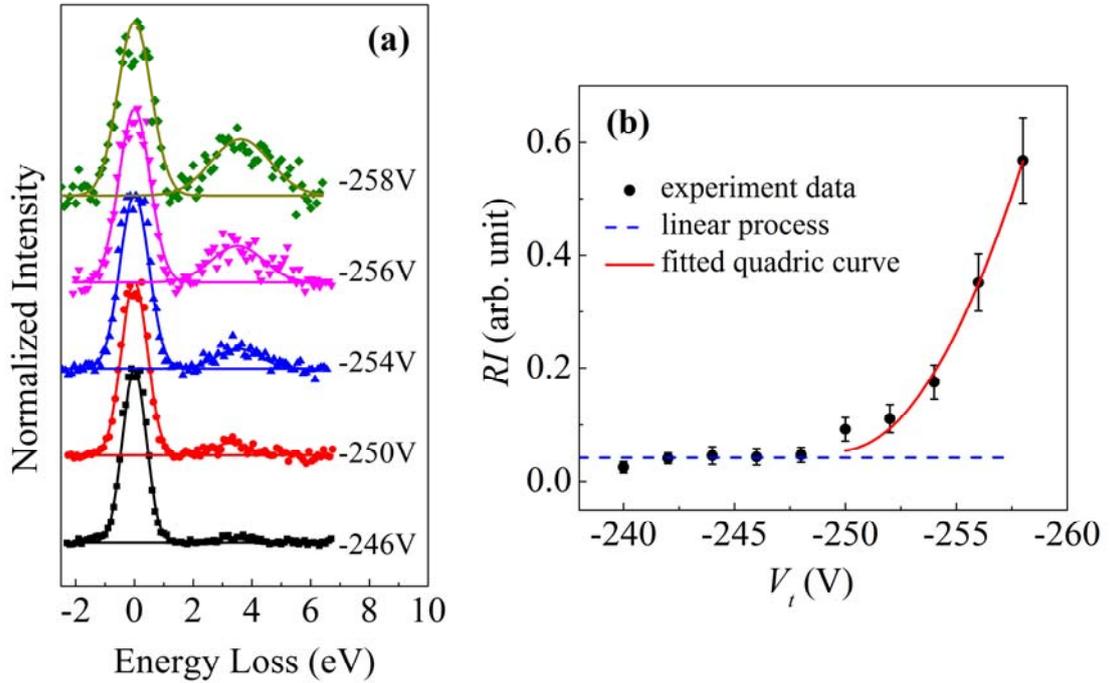

Figure 3: (a) Electron energy loss spectra obtained at five different tip biases with constant tip-sample distance, from which nonlinear processes of the SPP excitation are clearly observed. (b) Relative intensity (*RI*) of SPP energy loss peak versus tip voltage. The *RI* is almost keeping unchanged when the value of tip voltage is below 250V (blue dashed line), but increases with quadric dependence as the value of tip voltage is beyond 250V (red solid line).

Since the cross section of the elastic scattering is proportional to electron density $I_0 = kE^2$ ($k$ is the constant and $E$ electric field strength), the expression of *RI* should have the form of:

$$RI \propto \alpha + \beta I_0 = \alpha + \beta k E^2 \qquad (1)$$

where $\alpha$ and $\beta$ can be expressed as follows

$$\alpha \propto \left|\langle \psi_b | \hat{e}_1 \cdot \vec{D}_{ba} | \psi_a \rangle\right|^2 \quad (2)$$

$$\beta \propto \left|\sum_n \frac{\langle \psi_b | \hat{e}_2 \cdot \vec{D}_{bn} | \psi_n \rangle \langle \psi_n | \hat{e}_1 \cdot \vec{D}_{na} | \psi_a \rangle}{\varepsilon - \varepsilon_{na} + i\Gamma_n/2}\right|^2 \quad (3)$$

Here $\vec{D}_{ba}$, $\vec{D}_{bn}$ and $\vec{D}_{na}$ are the matrix elements of the multipole moment operator for electron scattering (it reduces to the dipolar term for SPP [11]), $\hat{e}_1$ and $\hat{e}_2$ are the unit vector of the electric field, $\varepsilon$ is the energy of the incident electron, $\varepsilon_{na}=\varepsilon_n-\varepsilon_a$ is the energy difference between an intermediate state $|\psi_n\rangle$ and the ground state $|\psi_a\rangle$, $\Gamma_n$ is the energy width of the state $|\psi_n\rangle$, and the sum over $n$ is over all possible intermediate states $|\psi_n\rangle$. The linear term $\alpha$ is associated with dipole scattering which was thought to be dominant for the generation of SPP.

As can be seen in Figure 3b, when the value of $V_t$ is below 250V, the *RI* is almost kept unchanged, indicating that the linear process dominates in this region. However, as the value of $V_t$ goes beyond 250V, nonlinear effects arises and the dependence of *RI* with $V_t$ fits the quadric curve quite well.

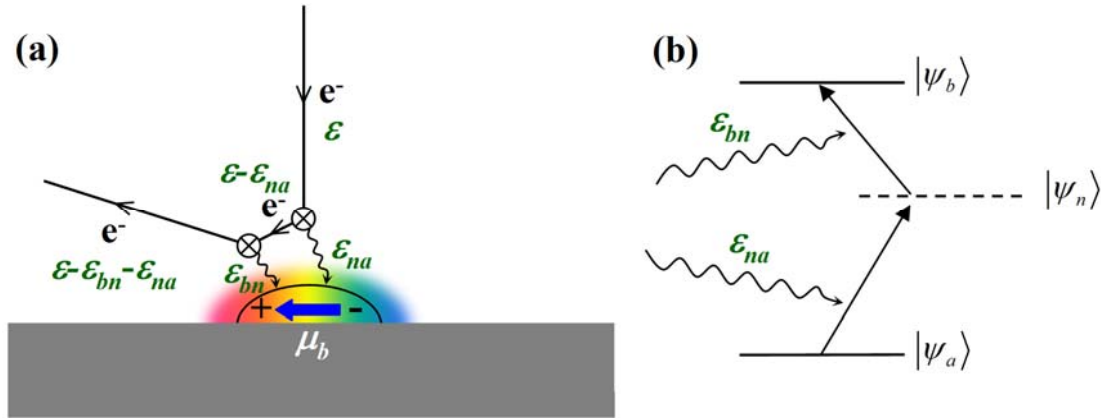

Figure 4. Schematics of the two-step scattering processes (a) and the corresponding energy level transition (b). An incident electron interacts with the nanostructure on the surface to excite it from the ground state $|\psi_a\rangle$ to an intermediate state $|\psi_n\rangle$ with the energy loss $\varepsilon_{na}$. During the interacting time interval, the electron can subsequently interact with the nanostructure again, exciting it to the final state $|\psi_b\rangle$ by losing energy $\varepsilon_{bn}$. The dipole moment of the SPP state $\mu_b$ is demonstrated as a blue arrow in figure (a).

The nonlinear coefficient $\beta$ could be contributed from two different physical processes due to either the interactions between the system and the subsequent two incident electrons or the two-step scattering with one incident electron. In our experimental setup, although the field can be significantly enhanced, the maximum field emission current at tip voltage -258V is only 1nA. One can estimate that the average number of electrons between the tip and sample is less than 0.1, strongly indicating that the interactions with the subsequent two incident electrons are negligible. The two-step scattering process experienced by one electron becomes the most possible mechanism, in which as illustrated in Figure 4, an incident electron interacts with the nanostructure on the surface to excite it from the ground state $|\psi_a\rangle$ to an intermediate state $|\psi_n\rangle$ with the energy loss $\varepsilon_{na}$. During the interacting time interval, the electron can subsequently interact with the nanostructure again, exciting it to the final state $|\psi_b\rangle$ by losing energy $\varepsilon_{bn}$. All intermediate states $|\psi_n\rangle$ can in principle contribute to this process. Obviously, the high density of intermediate states certainly helps to increase the value of the nonlinear coefficient, which is actually what Ag nanostructures can provide. The exceptional large transition and permanent dipolar moments of certain states distinguish themselves from others. One very important case is when only two states, the ground (a) and the final state (b), are involved. In this case, $\beta$ is controlled by one term

$$\left| \frac{\langle \psi_b | \hat{e}_2 \cdot \vec{D}_b | \psi_b \rangle \langle \psi_b | \hat{e}_1 \cdot \vec{D}_{ba} | \psi_a \rangle}{\varepsilon - \varepsilon_{ab} + i\Gamma_b/2} \right|^2 \propto \mu_b^2 \alpha \qquad (4)$$

Within the dipolar approximation, $\mu_b$ is the dipole moment of the final state. The excited SPP state of Ag nanostructure possesses extremely large dipole moment that makes it a perfect candidate to enhance the probability of nonlinear electron scattering. This particular scattering channel corresponds to a rather interesting physical picture: the incoming electron losses its energy to excite the Ag nanoclusters to its plasmon state, on which the outgoing electron is elastically scattered out.

It is noted that with a similar experimental setup, Palmer's group did not observed any nonlinear behavior for the SPP excitation of Ag surface [16,17]. This might largely due to the difference in samples. The samples used in their experiments were prepared by evaporating 200 nm Ag on HOPG that could result in a relatively flat Ag surface, whereas small islands of Ag nanostructures are clearly formed in our samples. It seems to suggest the nonlinear coefficient $\beta$ of the SPP states for the flat surface is still not large enough to generate the nonlinear electron scattering process. The use of Ag nanostructures which are known to enhance the intensity of SPP [2,3,14] might be the key for the current success.

In conclusion, we have shown in this study that the nonlinear electron scattering is a detectable process in electron energy loss spectroscopy when both local electric field strength and the nonlinear scattering coefficients of the sample are effectively enhanced, as nicely demonstrated by the measurements for SPP of Ag nanostructures in a scanning probe electron energy spectrometer. The experimental results can be explained by a generalized one-electron-two-step scattering mechanism, which suggests an essentially new physical process for the generation of SPP compared to the conventional dipole scattering mode. The large nonlinear coefficient of SPP might be the key to understand the recent experimental and theoretical findings that the tip-induced SPP can behave as coherent ultrafast high intensity electromagnetic field [18,19]. The observation of nonlinear electron scattering in combination with strong SPP may pave the way to design new high-sensitivity high-spatial-resolution spectroscopic techniques for nanoscience and nanotechnology.

This work was partly supported by the National Basic Research Program of China (Grant Nos. 2010CB923301, 2010CB923304), National Science Foundation of China (Grant Nos. 10404026, 20925311) and MOE '211' project.

[1] J. B. Pendry, Phys. Rev. Lett. **85**, 3966 (2000).

[2] S. Lal, S. Link, and N. J. Halas, Nature Photonics **1**, 641 (2007).

[3] S. Kawata, Y. Inouye, and P. Verma, Nature Photonics **3**, 388 (2009).

[4] P. E. Batson, Phys. Rev. Lett. **49**, 936 (1982).

[5] J. Nelayah, et al, Nature Phys. **3**, 348 (2007).

[6] S. Nie, and S. R. Emory, Science **275,** 1102 (1997).

[7] P. Andrew, and W. L. Barnes, Science **306**, 1002 (2004).

[8] S. A. Claridge, J. J. Schwartz, and P. S. Weiss, ACSNano **5**, 693 (2011).

[9] A. A. Lucas, and M. Šunjić, Phys. Rev. Lett. **26**, 229 (1971).

[10] M. Rocca, Surf. Sci. Rep. **22**, 1 (1995).

[11] A. L. Koh, K. Bao, I. Khan, W. E. Smith, G. Kothleitner, P. Nordlander, S. A. Maier, and D. W. McComb, ACSNano **3**, 3015 (2009).

[12] C. K. Xu, X. J. Chen, X. Zhou, Z. Wie, W. J. Liu, J. W. Li, J. F. Williams, and K. Z. Xu, Rev. Sci. Instrum. **80**, 103705 (2009).

[13] P. W. Barber, R. K. Chang, and H. Massoudi, Phys. Rev. Lett. **50**, 997 (1983).

[14] E. Bailoa, and V. Deckert, Chem. Soc. Rev. **37**, 921 (2008).

[15] L. Savio, L. Vattuone, and M. Rocca, Phys. Rev. B **67**, 045406 (2003).

[16] R. E. Palmer, B. J. Eves, F. Festy, and K. Svensson, Surf. Sci. **502-503**, 224 (2002).

[17] A. Pulisciano, S. J. Park, and R. E. Palmer, Appl. Phys. Lett. **93**, 213109 (2008).

[18] Z. C. Dong et al., Nat. Photonics **4**, 50 (2009).

[19] G. Tian, J.-C. Liu, and Y. Luo, Phys. Rev. Lett. **106**, 177401 (2011).